\documentclass[traditabstract]{aa}
\usepackage[varg]{txfonts}

\usepackage{graphicx}
\usepackage{natbib}
\usepackage{amsmath}
\usepackage{amssymb}
\usepackage{url}
\usepackage{color}

\newcommand{\xmm}[0]{XMM-\textit{Newton}}
\newcommand{\chan}[0]{Chandra}

\newcommand{\hd}[0]{HD~209458}
\newcommand{\hdb}[0]{HD~209458\,b}
\newcommand{\acen}[0]{$\alpha$~Cen}

\newcommand{\ctks}[0]{\mbox{cts\,ks$^{-1}$}}

\newcommand{\funit}[0]{\mbox{erg\,cm$^{-2}$\,s$^{-1}$}}
\newcommand{\lunit}[0]{\mbox{erg\,s$^{-1}$}}

\newcommand{\lya}[0]{Ly$\alpha$}

\begin{document}

\title{Coronal X-ray emission and planetary irradiation in \hd}
\author{S. Czesla \and M. Salz \and P.C. Schneider \and M. Mittag \and J.H.M.M.
Schmitt} \institute{Hamburger Sternwarte, Universit\"at Hamburg, Gojenbergsweg 112, 21029
Hamburg, Germany}
\date{Received ... / Accepted ... }

\abstract{\hd\ is one of the benchmark objects in the study of hot
Jupiter atmospheres and their evaporation through planetary winds. The
expansion of the planetary atmosphere is thought to be driven by high-energy EUV
and X-ray irradiation. We obtained new \chan\ HRC-I data, which unequivocally
show that \hd\ is an X-ray source. Combining these data with
archival \xmm\ observations, we find that the corona of \hd\ is
characterized by a temperature of about 1~MK and an emission measure of $7
\times 10^{49}$~cm$^{-3}$, yielding an X-ray luminosity of $1.6\times
10^{27}$~\lunit\ in the $0.124-2.48$~keV band. \hd\ is an inactive star with
a coronal temperature comparable to that of the
inactive Sun but a larger emission measure.
At this level of activity, the planetary high-energy emission is sufficient to
support mass-loss at a rate of a few times $10^{10}$~g\,s$^{-1}$.
}

\maketitle

\section{Introduction}

Planets orbiting their host stars at close proximity are thought to undergo
mass-loss through planetary winds driven by stellar X-ray and extreme
ultraviolet irradiation \citep[e.g.,][]{Salz2016a, Salz2016b}. This may even
affect the planetary evolution \citep{Sanz2010} and
thus, precise knowledge of the stellar high-energy spectrum is critical to
understand both the stars and their planets.

One of the key systems in which planetary mass-loss can be studied to date is \hd.
In fact, the G0 dwarf \hd\ is one of the optically brightest stars known to host
a transiting hot Jupiter, which orbits
the star every $3.5$~d at a distance of about $0.047$~AU \citep{Charbonneau2000,
Henry2000}.
\hdb\ was among the first planets for which
mass-loss was detected by means of enhanced in-transit absorption in the \lya\
line \citep{VidalMadjar2003}. Subsequently, further elements such as oxygen and
carbon were reported in the planetary wind \citep{VidalMadjar2004}. 

The distance to the \hd\
system was determined by GAIA to be $48.9 \pm 0.5$~pc \citep[DR1, ][]{Gaia2016a,
Gaia2016b}.
Using hydrogen Lyman\,$\alpha$ line observations,
\citet{Wood2005} determined an interstellar \ion{H}{i} column density of
$2.3\times 10^{18}$~cm$^{-2}$ toward \hd.
The star is slightly larger than the Sun
\citep[$1.2$~R$_{\odot}$,][]{Boyajian2015, delBurgo2016}, shows solar iron
abundance \citep[$\mbox{Fe/H}=0.02\pm 0.03$,][]{Santos2004}, and is a bona-fide
single \citep{Raghavan2006, Daemgen2009}.

According to its projected rotation velocity, the rotation period of \hd\ is
$14.4\pm 2.1$~d, assuming the system is aligned, which is consistent with
measurements of the Rossiter-McLaughlin effect \citep{Winn2005}.
\citet{SilvaValio2008} favor a somewhat shorter rotation period of $11.4$~d
based on in-transit starspot observations. According to \citet{Bonfanti2016},
the age of \hd\ is about $4.4$~Gyr, i.e., essentially the age of the Sun.
Yet, age estimates for \hd\ differ widely from around 2~Gyr, determined by
\citet{Maxted2015} using gyrochronology and isochrones,
to $6.5$~Gyr as reported by \citet{Boyajian2015} based on evolutionary
modeling, however, with a large uncertainty of nearly 3~Gyr.

Although \hd\ rotates at about twice the solar rate,
it is a rather inactive star. \citet{Henry2000} report a $\log R'_{HK}$ value of
$-4.93$. Between 2004 and the end of 2009,
\hd\ was observed in the context of the California
Planet Search \citep{Isaacson2010}. During this period,
the star showed a continuously low activity level with $\log R'_{HK}$
values ranging between $-4.930$ and $-4.974$, i.e., it remained close to the
minimum solar activity level of $\log R'_{HK} \approx -4.96$
\citep{Mamajek2008}. This is also consistent with the value of $-4.97$ reported by \citet{Knutson2010}.
From the coadded spectrum obtained between Aug. 2013 and Aug. 2014 by
the TIGRE telescope \citep{Mittag2011, Schmitt2014}, we measured a Mount-Wilson
S-index of $0.167\pm 0.006$ \citep[e.g.,][]{Mittag2016}. Applying the
calibration by \citet{Noyes1984}, this value can be converted into
$\log(R'_{HK}) = -4.91\pm 0.03$, comparable to the previously determined values.

Despite its critical role in planetary mass loss, 
the X-ray spectrum and luminosity of \hd\ remain controversial.
\citet{Sanz2011} determined an upper limit of $2.5\times 10^{26}$~\lunit\
for the X-ray luminosity of \hd\ ($0.12-2.48$~keV band), using \xmm\
data and assuming a coronal temperature of 2~MK.
The 3XMM catalog lists fluxes for a source with a
position about two arcseconds from that of \hd\ \citep[][]{Rosen2016}. Adopting
the distance of \hd, the reported flux yields an X-ray source with a luminosity
of $(6.7\pm 3.8) \times 10^{26}$~\lunit\ in
the $0.2-12$~keV band \citep{Louden2016}. The reconstruction
of the differential emission measure presented by \citet{Louden2016} yields an
X-ray luminosity of $(1.2 \pm 0.2)\times 10^{27}$~\lunit\ in the
$0.124-2.48$~keV band. The authors used the 3XMM catalog fluxes to constrain
their model. According to their analysis, the
mass-loss rate of the planet \hdb\ is $(3.8\pm 0.2)\times 10^{10}$~g\,s$^{-1}$.

Notably, all above estimates of the soft X-ray flux are fully or partly
based on the available \xmm\ data, which suffer from considerable background
contamination, complicating the analysis of a weak and soft source like \hd.  
Therefore, we obtained a new \chan\ HRC-I observation to constrain the soft
X-ray flux of \hd\ more stringently.

\section{Data analysis}
We carry out a detailed analysis of our new \chan\ HRC-I data, the available
archival \xmm\ observations and, finally, present a combined analysis of both
data sets.

\subsection{\chan\ HRC-I data}
\label{sec:chanHRCI}

\begin{figure}[h]
  \includegraphics[width=0.49\textwidth]{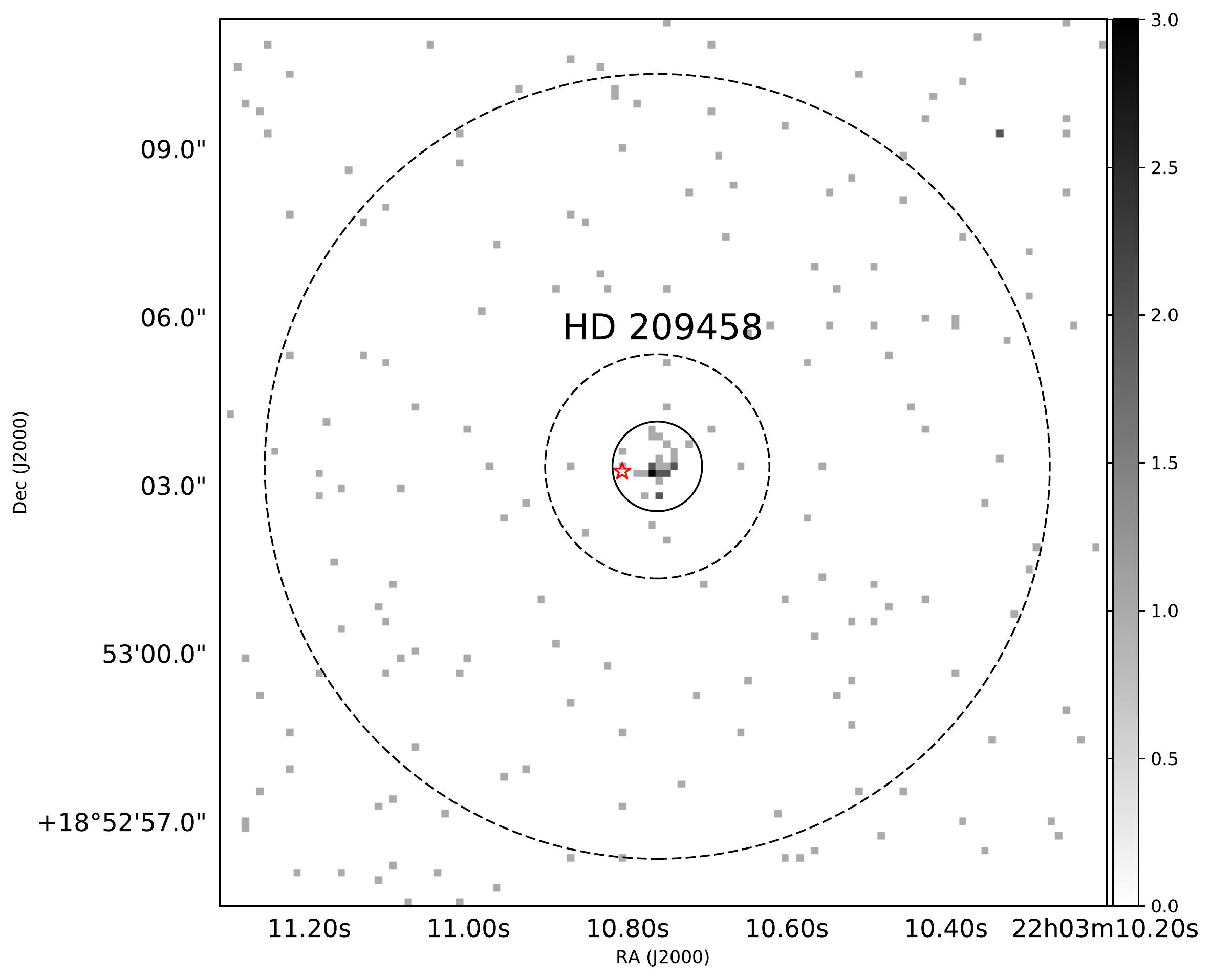}
  \caption{\chan\ HRC-I image of \hd\ with our circular source region, the
  background annulus (dashed), and a star symbol indicating the nominal,
  proper-motion corrected year 2016.5 position from GAIA.
  \label{fig:hrcImage}}
\end{figure}

\chan\ observed \hd\ on June 17, 2016, for $28.9$~ks using the HRC-I
(observation ID 16667). Our analysis is based on the pipeline-produced event
files and \texttt{CIAO} in version~4.8.
While the HRC-I does not provide any notable energy
resolution\footnote{See,
e.g., the report ``Spectral Response of the HRC-I'' by V.~Kashyap and
J.~Posson-Brown at
\url{http://cxc.cfa.harvard.edu/ccr/proceedings/05_proc/presentations/kashyap2/}},
its sensitivity extends to photon energies of $80$~eV, which allows to study
soft X-ray sources.
In Fig.~\ref{fig:hrcImage}, we show the HRC-I image of the surroundings of \hd.

We identify a clear X-ray source close to the nominal position of \hd\ (see
Fig.~\ref{fig:hrcImage}). 
As our source region, we use a circle with a radius of $0.8''$
also indicated in Fig.~\ref{fig:hrcImage}. Using the \texttt{CIAO} task
\texttt{psffrac}, we estimate that this region contains $93$\% of the photons
with an energy of $200$~eV.
We first center the source region on the visual position of the source and then
refine the position by determining the centroid of the photon distribution
(see Table~\ref{tab:positions}).
In the final source region, a total of $29$~counts are recorded.

The centroid position of the X-ray source is offset by $(0.6 \pm 0.2)''$ toward
the west with respect to the proper-motion corrected optical position of \hd\
determined by GAIA \citep[DR1, ][]{Gaia2016a, Gaia2016b}.
This offset is within the accuracy range of the Chandra aspect
solution\footnote{\url{http://cxc.harvard.edu/proposer/POG/html/chap5.html}},
so we are confident that the X-ray source is associated with the bright star
\hd\ (see also Sect.~\ref{sec:astrometry}).

\begin{figure}[h]
  \includegraphics[width=0.49\textwidth]{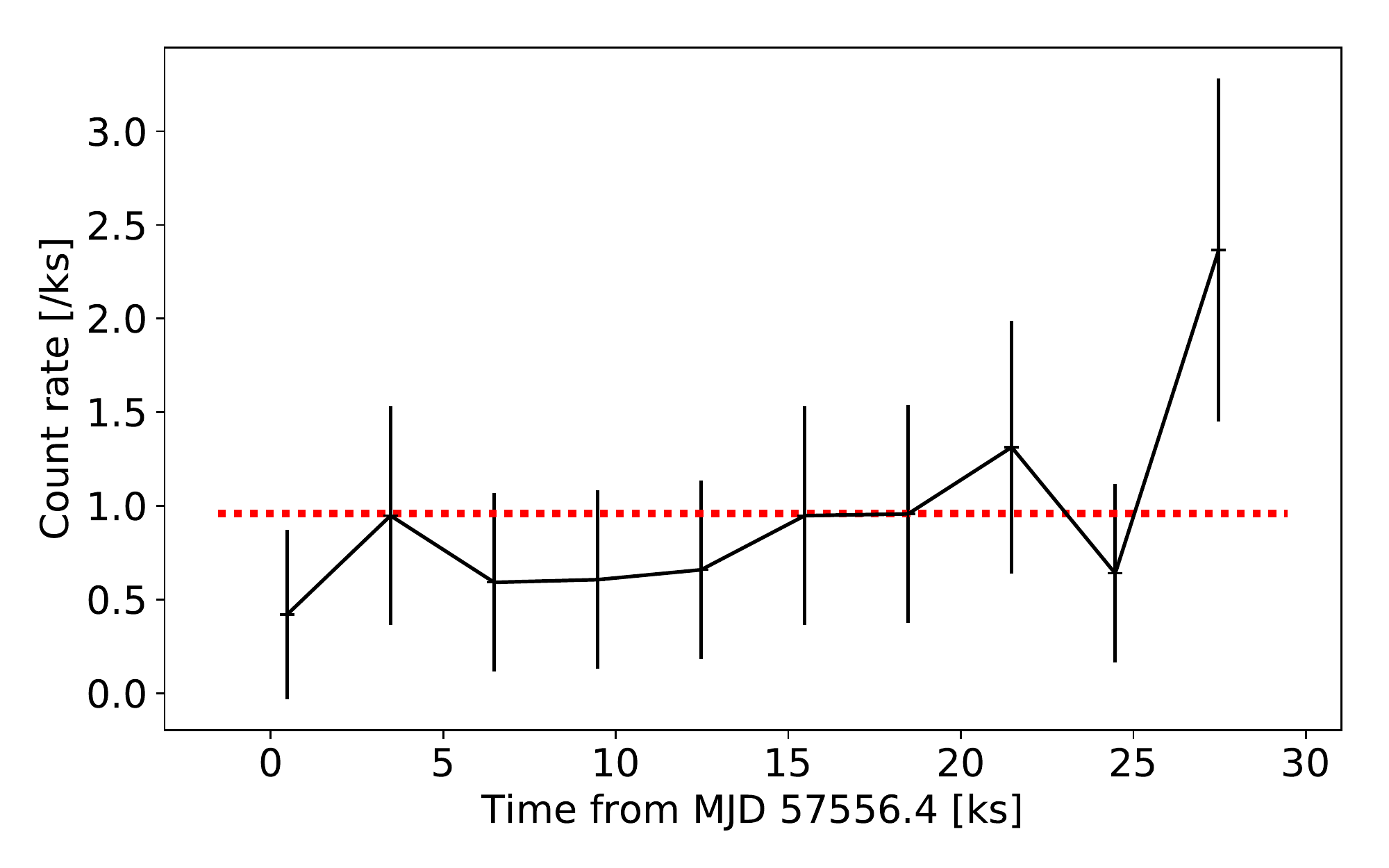}
  \caption{Background-subtracted HRC-I light curve of \hd\ with a temporal
  binning of $3$~ks along with the mean count rate (red, dotted).
  \label{fig:lc}}
\end{figure}

We estimate $1.4$ background counts in the source region by considering
an annulus with the same centroid and limiting radii of $2''$ and $7''$.
Given the exposure time of $28.9$~ks, this yields a net source
count rate of $(0.96 \pm 0.19)$~\ctks. The HRC-I light curve of \hd\ shows
no obvious variability such as a flare and is compatible with a constant flux
(see Fig.~\ref{fig:lc}).

\subsubsection{Refinement of HRC-I astrometry}
\label{sec:astrometry}

\begin{table}
\centering
\caption{Source coordinates determined from HRC-I data and obtained from GAIA
and the USNO-B1 catalog. 
\label{tab:positions}}
\begin{tabular}{l l l} \hline \hline
Source & RA [deg] & Dec [deg] \\ \hline
\multicolumn{3}{c}{\hd} \\ 
HRC-I & 330.794845 & 18.884262 \\
GAIA (2015) & 330.795018 & 18.884244 \\
\hline
\multicolumn{3}{c}{Reference source} \\
HRC-I & 330.801708 & 18.943287 \\
USNO-B1 & 330.801887 & 18.943328 \\ \hline
\end{tabular}
\end{table}

The closest reference X-ray source with respect to the image center we
could identify, using the celldetect algorithm, is located 3.6 0 nearly
due north of HD 209458.
--> The closest reference X-ray source is located 3.6' nearly due north of
HD 209458. (Verschachtelter Satzbau)

Although we are confident that the small positional offset between the X-ray
source and the optical position of \hd\ is instrumental, this can be
validated by finding other X-ray sources in the field with known optical
counterparts. The closest reference X-ray source 
we could identify, using the \texttt{celldetect} algorithm, is located $3.6'$
nearly due north of \hd. The centroid position of the source with about
$40$~registered net counts, corresponds to a faint red object in the USNO-B1
catalog \citep[source ID \mbox{089-0583168}, R$_2 = 19.9$~mag,][]{Monet2003}.
This source is located $0.6''$ nearly due east of the X-ray position, which
supports the contention that the offset seen for \hd\ is in fact due to a
slight error in the aspect solution.

\subsection{\xmm{} data}

\xmm\ observed \hd{} in the years 2000 and 2006
(see Table~\ref{tab:obs} for details)\footnote{An additional observation in
2003 (Obs.
ID 0148200101) is not scientifically usable because of a filter-wheel failure.}.
In the year 2000 observation, optical loading caused by the use of the thin
filter, leaves only the MOS2 data usable for our
analysis.

We reduced the data using the Scientific Analysis
System (SAS) in version 16.0.0\footnote{``Users Guide to the XMM-Newton Science
Analysis System'', Issue 12.0, 2016 (ESA: XMM-Newton SOC).}.
In the reduction, we used standard reduction recipes. We used the
\texttt{epreject} task to improve the pn energy scale particularly at low
photons energies.
The high-energy background of the observations is well behaved with only the
2006 pn observation showing a weak, approximately 2~ks long background flare,
which is excluded from our analysis.

\begin{table}
  \caption{XMM-Newton observations of \hd{}}             
  \label{tab:obs}      
  \centering          
  \begin{tabular}{l @{\;\;} l l @{\;\;\;} l @{\;\;} l }
    \hline\hline  
    \vspace{-9pt}\\      
      Obs. ID/ & Camera & Filter & Duration & Livetime \\
      Date     &        &        & [ks] & [ks] \\
    \vspace{-9pt}\\ 
    \hline
    \vspace{-9pt}\\ 
      0130920101  & MOS1 & thin\tablefootmark{a}  & 17.3 \\  
      2000-Nov-29 & MOS2 & thick & 18.1 & 16.2 \\  
                  & pn   & thin\tablefootmark{a}  & 16.4
                 \\ \hline 
    \vspace{-9pt}\\
      0404450101 & MOS1 & medium & 33.3 & 32.8 \\  
      2006-Nov-14 & MOS2 & medium & 33.3 & 32.8 \\  
                  & pn   & medium & 31.6 & 26.4 \\  
    \vspace{-9pt}\\ 
    \hline                  
  \end{tabular}
  \tablefoot{
    \tablefoottext{a}{Corrupted by optical loading.}
  }
\end{table}

\subsubsection{Detection experiment}

To test whether X-ray emission of \hd\ can be detected in the
\xmm\ data, we carried out a detection experiment.
For all scientifically usable exposures, we defined a
circular source region with a radius of 15\arcsec, centered on
the source position as determined from the Optical Monitor images; none of these
is more than $2.5"$ from the nominal optical position.
We then extracted the counts in these source regions in a
soft ($0.15-0.5$~keV) and a hard spectral band ($0.5-1.5$~keV), which together
comprise the bulk of X-ray emission from rather inactive stars such as \hd, detectable
with \xmm's EPIC instruments. While no emission at higher photon energies is
expected, there may be substantial emission at lower energies, not covered by \xmm. 

\begin{figure*}
  \includegraphics[width=0.49\textwidth]{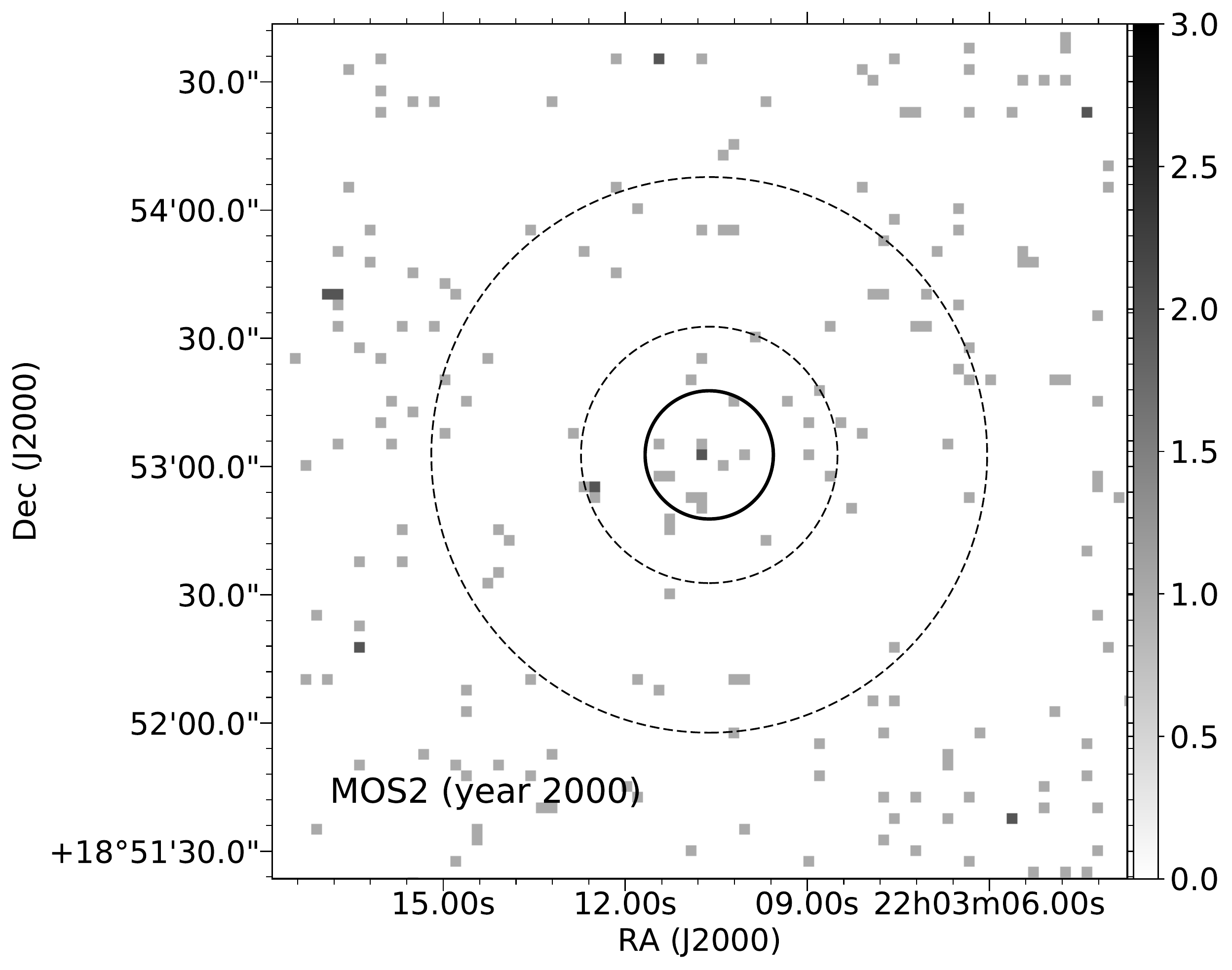} 
  \includegraphics[width=0.49\textwidth]{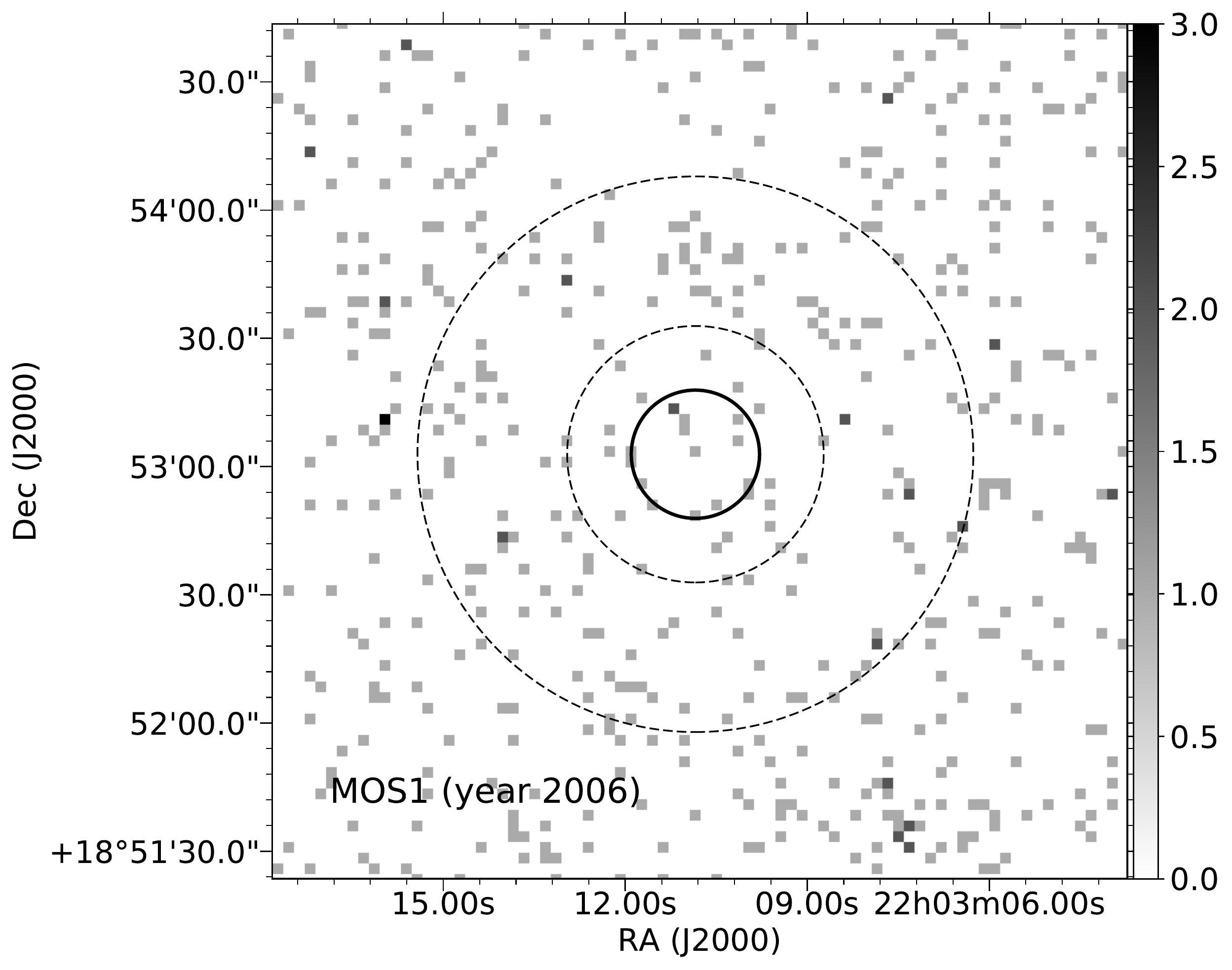} \\
  \includegraphics[width=0.49\textwidth]{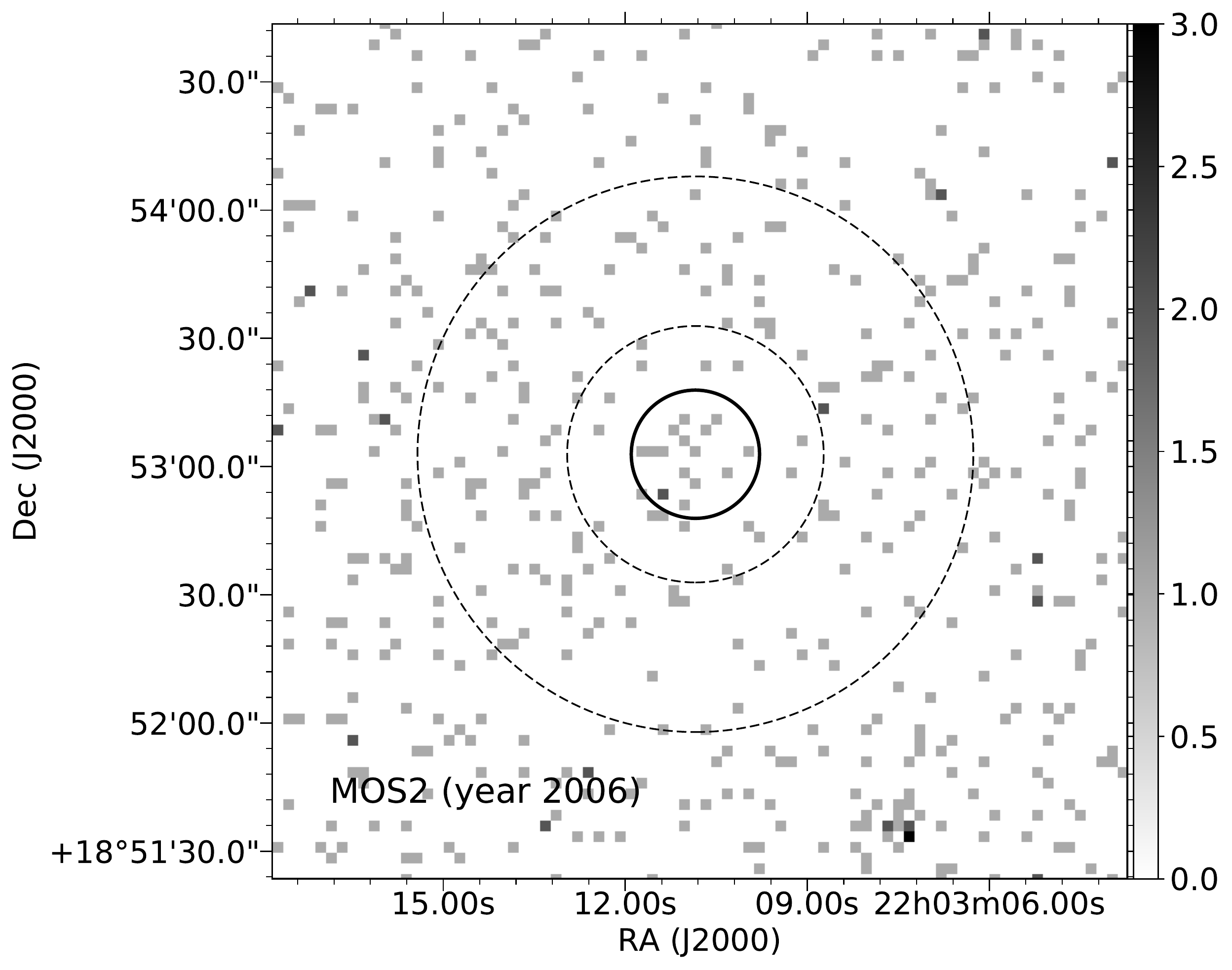}
  \includegraphics[width=0.49\textwidth]{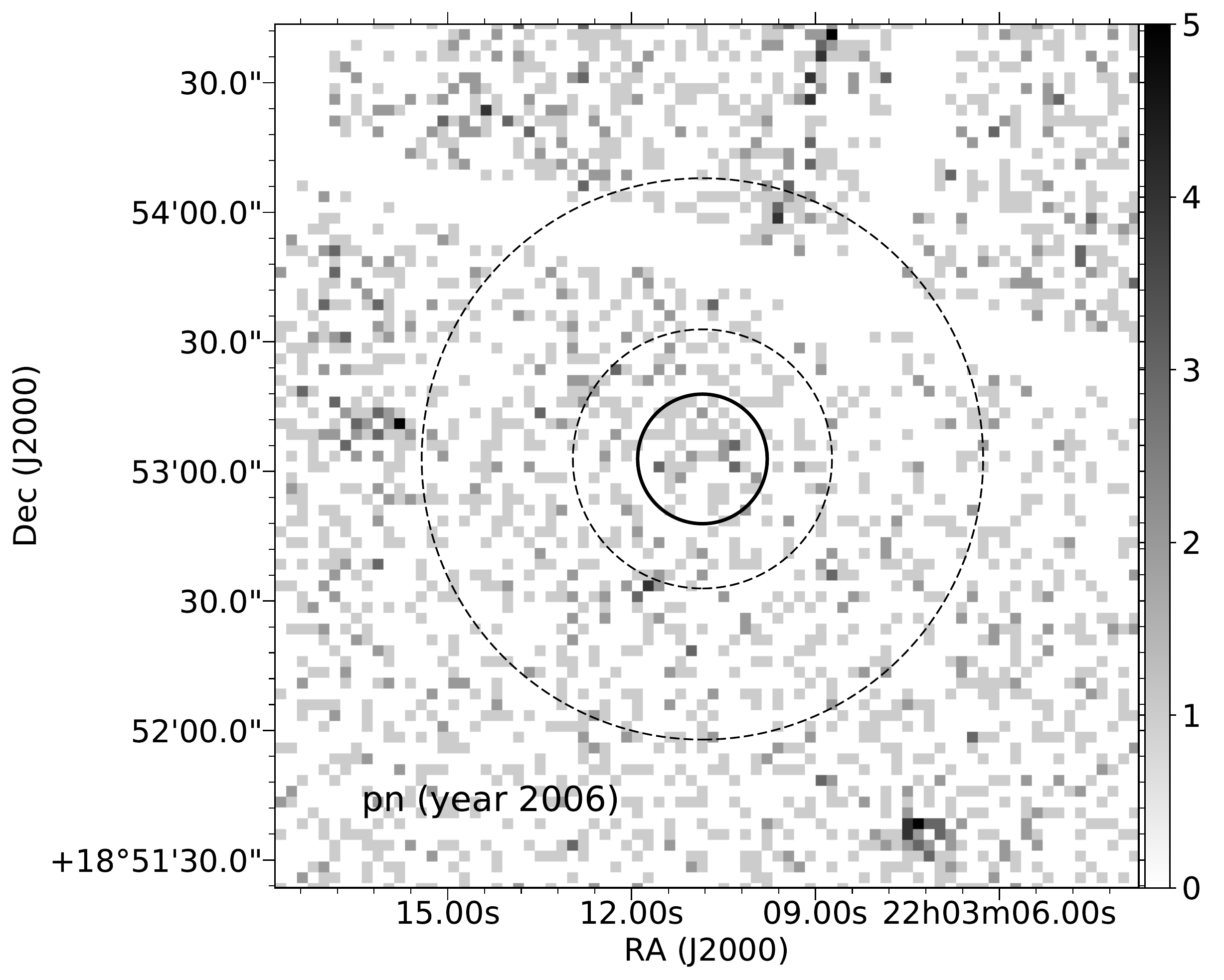}
  \caption{\xmm\ X-ray images of \hd\ and its surroundings in the $0.15-1.5$~keV
  band (EPIC detector and year of observation are indicated in the panels).
  The solid circle indicates the 15'' source region and the dashed annulus the background region.
  \label{fig:xmmim}}
\end{figure*}

After visual inspection of the X-ray images, we defined an annulus-shaped
background region with an inner radius of $30$\arcsec\ and an outer radius of
$65$\arcsec\ (see Fig.~\ref{fig:xmmim}). This background region does not cover
obvious X-ray sources in the vicinity of \hd; also the \chan\ HRC-I image does
not show any sources in this region (see Sect.~\ref{sec:astrometry}).
Only for the 2006 pn observation, the background annulus covers the
chip edges, which does not appear to introduce any particular artifacts. Generally, the
background in the vicinity of \hd\ seems to be inhomogeneous so that differently
placed circular reference regions yield different rates. Ignorant of the actual
source-region background, we consider the annulus to yield an appropriate
estimate.
The detected source and background counts in the soft and hard band are listed
in Table~\ref{tab:Counts}. Along with these numbers we list the number of net
counts, $n_{\rm net}$, calculated according to
\begin{equation}
  n_{\rm net} = s - f_a\, b \;\;\;\mbox{and}\;\;\; \sigma_{\rm net} = \sqrt{s +
  f_a^2 b} \; ,
\end{equation}
where $s$ and $b$ are the number of photons detected in the source and
background region and $f_a$ the ratio of their areas. Of course, negative values
are not physically admissible, and a more stringent statistical treatment is
applied in Sect.~\ref{sec:combinedAnalysis}.

\begin{table}
\centering
\caption{Counts detected in the 15\arcsec\ source region (SRC) and the
$30-65$\arcsec\ background annulus (BGA) in the individual EPIC
instruments in the years 2000 and 2006.
\label{tab:Counts}}
\begin{tabular}{l l r r r} \hline \hline
Instr. & Band & SRC & BGA & Net counts \\ \hline
\multicolumn{5}{c}{Year 2000} \\
MOS2 &  soft &   3  & 18 & $1.75 \pm 1.76$ \\
MOS2 &  hard &   8  & 18 & $6.75 \pm  2.84$\\
\multicolumn{5}{c}{Year 2006} \\
 MOS1 & soft &  8  & 52 & $4.45 \pm 2.87$ \\
 MOS2 & soft &  13 & 42 & $10.03 \pm  3.63$ \\
 pn  & soft &  44 & 312 & $19.94  \pm 6.77$ \\
  MOS1 & hard &   4 & 87 & $-1.94 \pm 2.10$ \\
 MOS2  & hard &   3 & 93 & $-3.58  \pm 1.86 $ \\
  pn  & hard &   16 & 183& $1.89 \pm 4.13$ \\ \hline
\end{tabular}
\end{table}

In the soft band, all individual observations yield a surplus of events in the
source region compared to the background. While not overly significant
individually, all soft-band results may be combined to yield a total of $36.2 \
\pm 8.4$~counts, which yields a significant detection in the
soft band.
In the year 2000 MOS2 observation, we
even find a slightly larger excess in the hard band than in the soft band, which
is not the case for the year 2006 observations, where only the pn yields a
marginally positive net count rate in the hard band.

\subsection{EPIC pn spectral analysis for year 2006}

\begin{figure}[hb]
    \includegraphics[width=0.49\textwidth]{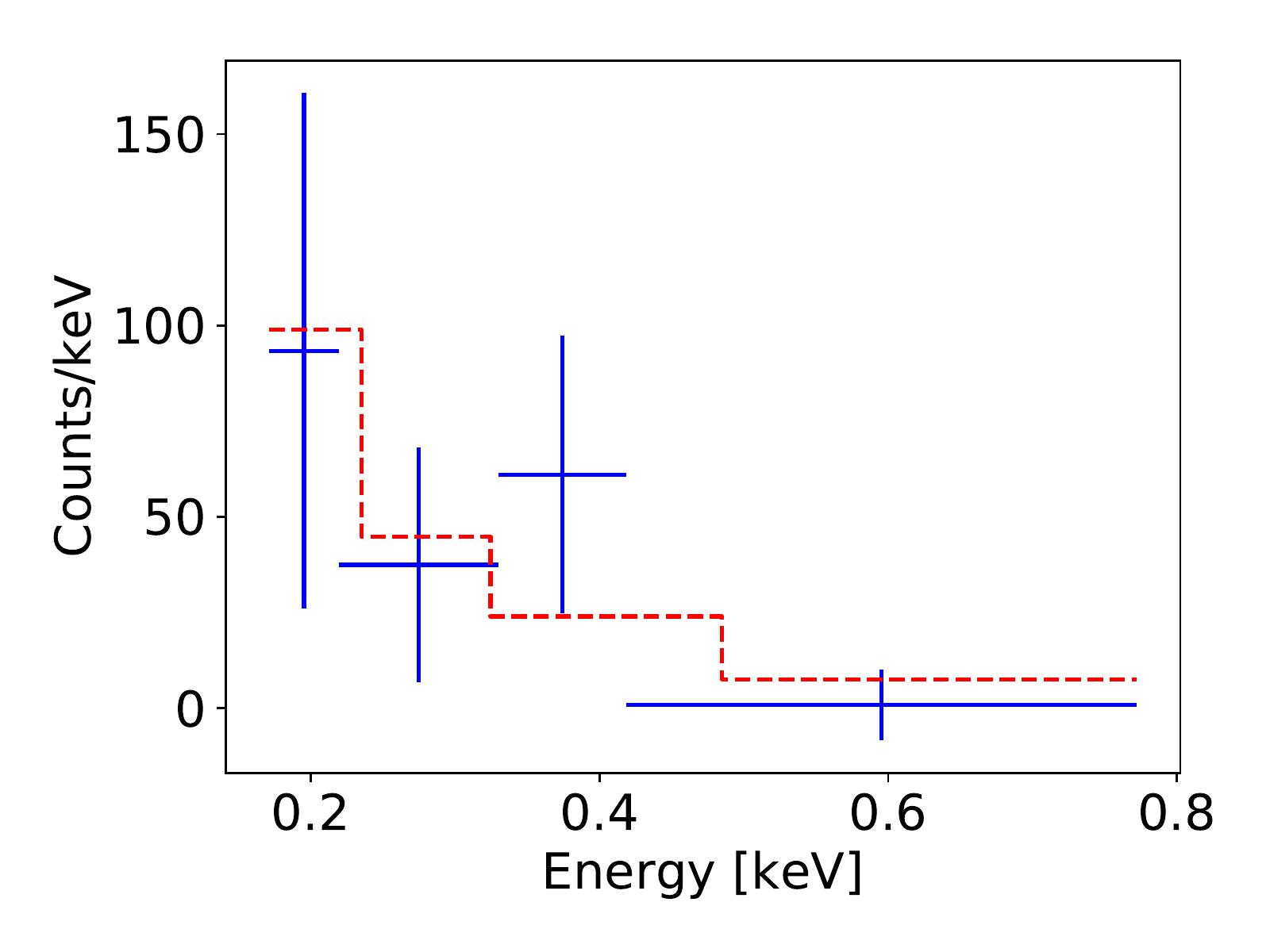}
  \caption{\xmm\ pn spectrum in 2006 with ten counts per bin (solid blue) and
  best-fit model (red, dashed).
  \label{fig:pnspec2006}}
\end{figure}

The year 2006 observation by \xmm's pn camera provides sufficient photons for a
coarse spectral analysis, which we carried out using XSPEC
\citep[version 12.9.1,][]{Arnaud1996}.
In Fig.~\ref{fig:pnspec2006}, we show the pn spectrum along
with the best-fit one-temperature thermal APEC model \citep{Foster2012}. In the
modeling, the interstellar hydrogen column density remained fixed at the value of
$2.3\times 10^{18}$~cm$^{-2}$ determined by \citet{Wood2005} and the
metallicity is taken to be solar.

The model yields a best-fit temperature of $0.08_{-0.04}^{+0.06}$~keV with the
error indicating the $90$\% confidence limits. The best-fit emission measure is
$0.9\times 10^{50}$~cm$^{-3}$, which remains rather loosely constrained. In
fact, the $90$\% confidence limit only provides an upper limit of
$21\times 10^{50}$~cm$^{-3}$, which is essentially due to a degeneracy between
temperature and emission measure in this range.

\section{Combined HRC-I and \xmm\ analysis}
\label{sec:combinedAnalysis}

At our disposal we find a rather heterogeneous data set. A total
of three observational epochs in the years 2000, 2006, and 2016 are available,
and the data of no two epochs were obtained with the same
instrumental setup.

While the \chan\ HRC-I observation yields a clear detection of the X-ray
emission, it provides no spectral resolution in itself. The latter can be
provided by \xmm, the observations of which however, suffer from comparably
high background contamination, allowing only a detection of somewhat debatable
quality. Nonetheless, all observations provide information on the X-ray emission
of \hd.

Therefore, we carry out a combined analysis of all three observational epochs.
Although we are aware that also the X-ray emission of inactive stars like \hd\
and the Sun can change considerably on the timescale of years
\citep[e.g.,][]{Peres2000, Robrade2012, Ayres2014}, the hypothesis in the
following will be that all observations shall be explained by the same physical
model with identical coronal properties. Specifically,
we assume an absorbed,
single-temperature thermal APEC model with the interstellar
absorption column fixed to the known value. Thus, we end up
with two free parameters, viz., the plasma temperature, $T$, and the emission
measure, $EM$. Besides spectral variability, the cross-calibration of the
instruments introduces an additional uncertainty
\citep[e.g.,][]{Poppenhaeger2009, Robrade2012}.

We proceed by using \texttt{PyXSPEC} to derive model count rates as a function
of plasma temperature and emission measure for all available observations and
the soft and hard spectral bands. The corresponding response files and auxiliary
response files were obtained using \texttt{CIAO} and \texttt{SAS} routines,
respectively.

We use the source and background counts obtained for the
\chan\ HRC-I (Sect.~\ref{sec:chanHRCI}) and the rates obtained for
\xmm\ in the soft and hard spectral bands (Table~\ref{tab:Counts}). Assuming
uniform priors on the emission measure ($EM$) and plasma temperature, $T$, our
posterior reads
\begin{equation}
  p(EM, T|I) \sim \prod_{i} \frac{P(b_i|\lambda_{b_i}\,e_i)}{\lambda_{b_i}}
  P(s_i|\left(f_{a_i}\lambda_{b_i} + \lambda_{s_i}(EM,T)\right)\,e_i )
\end{equation}
where the index $i$ is meant to run over all observations and the soft and hard
spectral bands with the exception of the HRC-I observation, where no spectral
information is available. By $\lambda_{s_i}(EM,T)$, we refer to the model source
count rate as a function of emission measure and temperature and $s_i$ and $b_i$
denote the number of photons detected in the source and background region
referred to by the observation and spectral band indicated by $i$. We use an
inverse-rate prior on the background rate \citep[$\lambda_b$,][]{Jaynes1968}.
The factor $f_{a_i}$ is the area scaling between source and background and $I$ summarizes all
information used in the posterior.

\begin{figure}
    \includegraphics[width=0.49\textwidth]{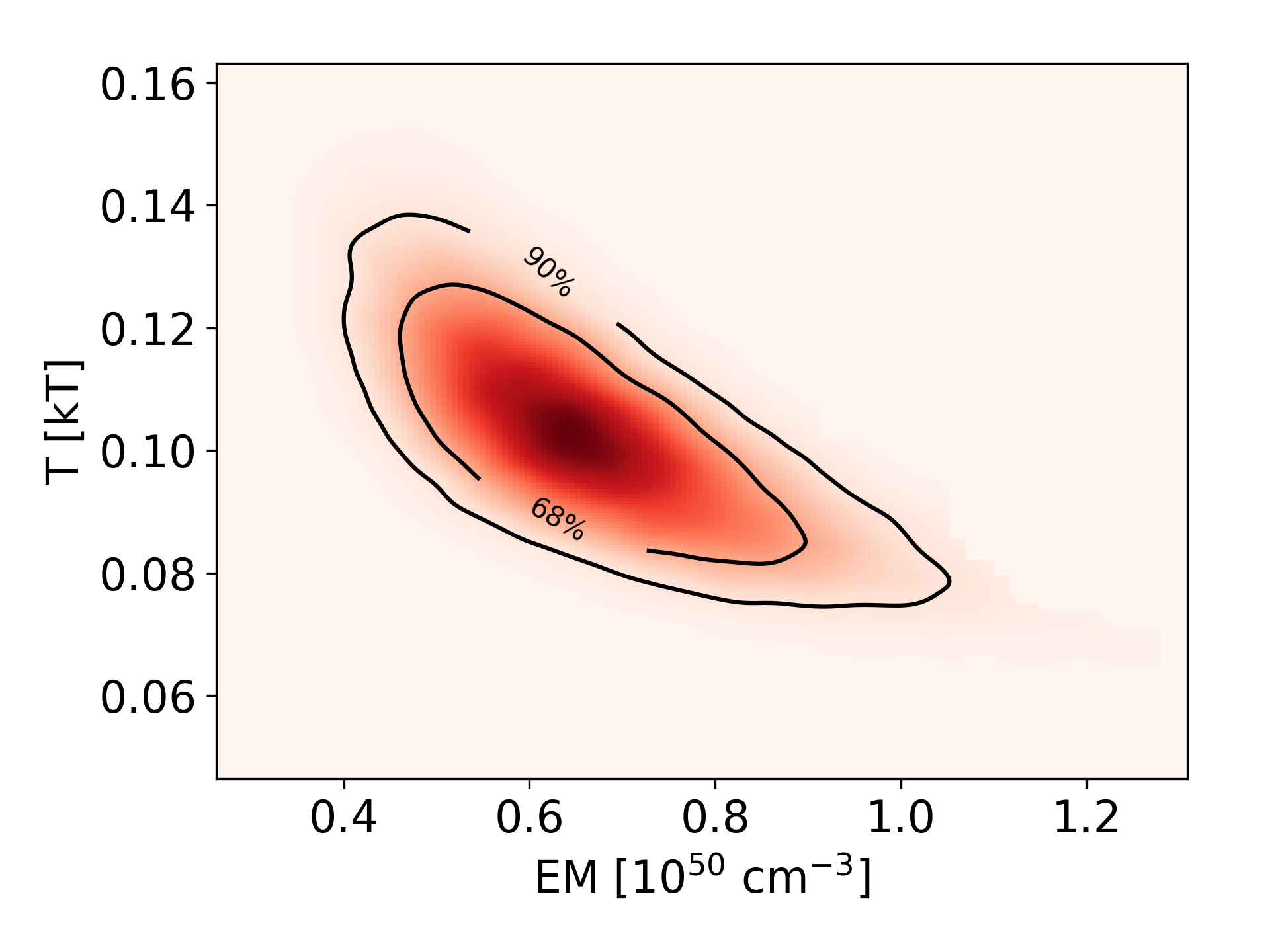}
  \caption{Posterior probability distribution for emission measure and coronal
  temperature from combined analysis of all available observations of \hd.
  \label{fig:postCombined}}
\end{figure}

In Fig.~\ref{fig:postCombined} we show the posterior density along with the $68$\% and
$90$\% highest probability density credibility regions. The marginal distributions
yield a temperature estimate of $0.10\,(0.07,0.13)$~keV (or $\log_{10}(T\,
\mbox{[K]}) = 6.1\,(5.9, 6.2)$) and an emission measure of $0.7\,(0.4,
0.9)\times 10^{50}$~cm$^{-3}$, which
corresponds to an X-ray luminosity of $8.4\,(5.8, 11) \times
10^{26}$~\lunit\ in the $0.2-12$~\AA\ band.
In the
$0.124-2.48$~keV band considered by \citet{Louden2016}, we obtain an X-ray
luminosity of $1.6\,(1.2,2)\times 10^{27}$~\lunit, consistent with their
results. In this band, the level of irradiation at the distance of the planet
is $255\,(190,320)$~\funit.

\subsection{Variability}

\begin{figure}[h]
  \includegraphics[width=0.49\textwidth]{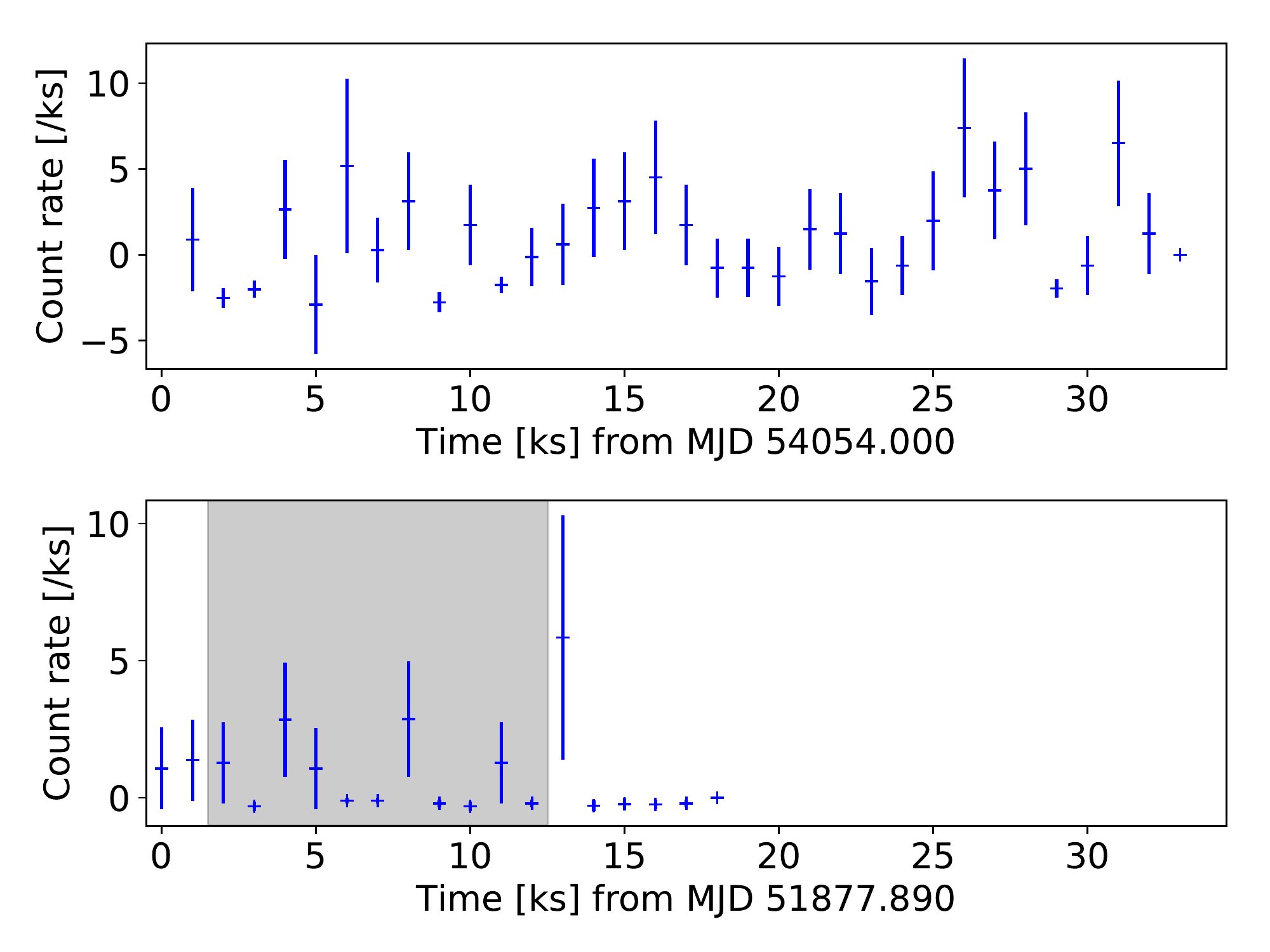}
  \caption{Background-subtracted $0.15-1.5$~keV \xmm\ light curves with a
  binning of $1$~ks. The top panel shows the 2006 pn light curve and the lower
  panel the 2000 MOS2 light curve. The shaded area in the bottom panel indicates
  the transit duration.
  \label{fig:xmmlcs}}
\end{figure}

Neither the \chan\ HRC-I nor \xmm\ observations show significant short-term
variability such as a flaring (see Sect.~\ref{sec:chanHRCI} and
Fig.~\ref{fig:xmmlcs}). The year 2000 \xmm\ observation covers a planetary
transit, which remains without noticeable effect due to the low count rate.
Nonetheless, the X-ray emission of \hd\ may be variable on the timescales of
years.
In Table~\ref{tab:CombinedCounts}, we show the expected number of source counts
along with the $90$\% credibility limits derived from the combined analysis of
the observations; for convenience and reference, the estimated
net counts are reproduced.

\begin{table}
\centering
\caption{Estimated net counts (replicated from Table~\ref{tab:Counts}) and
expectation based on the combined modeling. 
\label{tab:CombinedCounts}}
\begin{tabular}{l l r r } \hline \hline
Instr. & Band & Net counts & Model counts \\ \hline
\multicolumn{4}{c}{Year 2000 (\xmm)} \\
MOS2 &  soft & $1.75 \pm 1.76$ & 1.17\,(0.8,1.6) \\
MOS2 &  hard & $6.75 \pm  2.84$ & 0.4\,(0.1,0.6) \\
\multicolumn{4}{c}{Year 2006 (\xmm)} \\
 MOS1 & soft &  $4.45 \pm 2.87$ & 3.2\,(2.1, 4.3) \\
 MOS2 & soft &  $10.03 \pm  3.63$ & 3.5\,(2.4, 4.6) \\
 pn  & soft &  $19.94  \pm 6.77$ & 23.8\,(16.6, 31) \\
  MOS1 & hard &  $-1.94 \pm 2.10$ & 1\,(0.2, 1.8) \\
 MOS2  & hard &  $-3.58  \pm 1.86 $ & 1\,(0.2, 1.9) \\
  pn  & hard &  $1.89 \pm 4.13$ & 5.8\,(1.2,10.5) \\
\multicolumn{4}{c}{Year 2016 (\chan)} \\
HRC-I & $-$ & $27.6\pm 5.4$ & 29.4\,(20.8, 37.7) \\ \hline
\end{tabular}
\end{table}

While we generally find good agreement, the year~2000 observation with
\xmm's MOS2 shows a conspicuous excess of hard-band photons, which is not seen
during the year~2006 observation and not accounted for by the model.
This may, indeed, indicate a change in the coronal state of \hd\ between
the years 2000 and 2006. In the year 2000 observation, a hotter component may
have been present similar to the maximum state of the Sun.
However, we consider the supporting evidence too slim to draw a firm
conclusion.

\section{Comparison to other stars}

According to \citet{Schmitt1997},
the minimum stellar X-ray surface flux of F, G, and K type stars in the solar
vicinity is \mbox{$\approx\!10^4$~\funit{}}.
For \hd\ this translates into a minimum X-ray luminosity of $8\times
10^{26}$~\lunit\ in the ROSAT band, which is comparable to the value found in
our analysis. Therefore, the activity level of \hd\ is certainly at the lower
end of the observed range. To check whether it may even be exceptional, e.g.,
due to its hot Jupiter, we now
put the X-ray emission of \hd\ in the context of other low-activity stars.

\subsection{The $\alpha$~Centauri system}

The $\alpha$~Centauri system consists of a G2~V A~component, a K1~V
B~component, and the red dwarf Proxima~Centauri, orbiting the AB system in a
wide orbit. With an age of $6.5\pm0.3$~Gyr, the system
is older than \hd\ \citep{Eggenberger2004}. The A and B components have
extensively been monitored using \xmm\ \citep{Robrade2012} and the \chan\
HRC-I. A detailed analysis of the \chan\ data was presented by
\citet{Ayres2014}.
Specifically, the problem of obtaining factors to convert HRC-I
count rates into energy fluxes has been discussed by the author. 

According to \citet{Ayres2014}, the B~component shows an $8.1\pm0.2$~yr
solar-like coronal activity cycle. The A~component may have remained stuck in a
Maunder-Minimum-like state until 2005, but appears to show an elevation in
activity levels afterward, which may be indicative of a longer coronal cycle.

In the low activity state of \acen~A, \citet{Ayres2014} finds that its
corona is dominated by plasma at
temperatures slightly above $1$~MK, which is comparable to our result for
\hd. The two coronae may therefore, be similar. 
Although the intrinsic energy resolution of the HRC-I is
insufficient for a meaningful spectral analysis, some information on hardness is
available. We test whether the HRC-I spectrum of \hd\ is consistent with that of
the \acen\ sources, by comparing the photon energy distributions as given by the
\texttt{PI}\footnote{Pulse Invariant} channel. To this end,
we downloaded the 2005 HRC-I data of $\alpha$~Cen from the \chan\ archive and
extracted the photons corresponding to the $\alpha$~Cen A and B components; both
are strong HRC-I sources so that background is negligible. In this period, the
B component was observed close to its activity maximum, while the activity level
of the A component was low.
Using a Kolmogorov-Smirnov test, we compared the nominal photon energy
distributions of the \acen\ components with that of \hd. The resulting p-values
are $0.77$ for the A~component and $0.07$ for the B component. The test thus
also provides no evidence against the hypothesis that the \acen~A in its
low-activity state observed in 2005 and \hd\ show the same HRC-I spectrum.

Using the energy
conversion factor provided by \citet{Ayres2014} (Eq.~2), we estimate an X-ray
luminosity of $(9 \pm 2)\times 10^{26}$~\lunit\ for \hd\ in the $0.2-2$~keV band
based on the HRC-I count rate alone. Again, this is entirely consistent with our previous
analysis.

Compared to \acen~A in 2005, the X-ray luminosity of \hd\ is about
a factor of three higher. Both stars have similar radii of
$1.224$~R$_{\odot}$ and $1.155$~R$_{\odot}$ and thus surface areas
\citep{Kervella2003, Torres2008}, but \acen~A rotates with a period of $22.5\pm
5.9$~d, i.e., more slowly than \hd\ \citep{Bazot2007}.
Although this is consistent with a lower activity level, both \acen~A and \hd\
may of course show large, potentially cyclic variation in their coronal X-ray emission.
While \chan's HRC-I is not well-suited to resolve spectral variations in
the coronal emission of \hd, changes in its luminosity could well
be resolved by future observations.

\subsection{51~Peg}
Similar to \hd,
the G5V star 51~Peg also hosts a hot Jupiter \citep{Mayor1995}. This star
shows a sufficiently low activity level as to qualify
as a Maunder-Minimum candidate \citep{Poppenhaeger2009}. \citet{Baliunas1995}
present a Mount-Wilson S-index time series of 51~Peg ranging from 1977 to 1991, during
which the star remained relatively stable with a mean S-index of $0.149$.
Our more recent measurements with TIGRE yield a consistent S-index of
$0.152\pm 0.002$ \citep{Mittag2016}, providing no indication for a change in
the activity state of 51~Peg.

\citet{Poppenhaeger2009} studied the X-ray emission of 51~Peg and, among others,
analyzed a \chan\ HRC-I pointing. Form their analysis, the authors conclude that
the coronal temperature of 51~Peg is $\lesssim 1$~MK.
Scaling the HRC-I count rate of 51~Peg reported by \citet{Poppenhaeger2009} to
the distance of \hd, we calculate a distance-scaled count rate of
$0.42\pm0.1$~ct\,ks$^{-1}$. This implies that 51~Peg is less luminous in X-rays
than \hd\ by about a factor of two, which is also consistent with the
low coronal temperature derived by \citet{Poppenhaeger2009}.

\subsection{The Sun}

Due to a lack of appropriate instrumentation, continuous long-term studies
of the broad-band soft solar X-ray emission remain difficult. \citet{Peres2000}
studied the Sun as an X-ray star, using
data taken by Yohkoh's soft X-ray telescope (SXT), which remained, however, less
sensitive at soft photon energies \citep{Tsuneta1991}, increasing the
uncertainty at the soft end of the spectrum.

\citet{Peres2000} simulated solar
ROSAT spectra corresponding to the minimum and maximum of the solar cycle.
During minimum, they found that the ROSAT spectrum can be fitted with a single
1~MK component with an emission measure of $1.25\times 10^{49}$~cm$^{-3}$.
In solar maximum, a hotter ($2.6$~MK) component
appears and the cooler 1~MK component increases in emission measure by about on
order of magnitude. 
According to \citet{Peres2000}, the solar X-ray luminosity in the
ROSAT/PSPC band (here $0.1-3$~keV) varies between about
$0.23\times 10^{27}$~\lunit\  and $4.4 \times 10^{27}$~\lunit\ 
between solar minimum and maximum.

These values are quite similar to those we find for \hd. According to our
combined modeling of all available observations, the stellar corona is
characterized by a temperature of around $1$~MK and an
emission measure of $(4-9) \times 10^{49}$~cm$^{-3}$.
The coronal temperature of \hd\ is thus comparable to that observed
during solar minimum, but the emission measure corresponds to a value somewhere
between solar minimum and maximum. Only a fraction of this can be attributed
to the larger radius, leading to a $44$\% increase in the surface area
compared to the Sun. During the year 2000 observation, \hd\ may have shown a more active
state perhaps similar to a solar maximum. However, the evidence here presented
remains insufficient to draw such a conclusion.

\section{Conclusion}

We analyzed \chan\ HRC-I and \xmm\ data to determine the X-ray properties of
\hd. Our HRC-I data unequivocally establish \hd\ as an X-ray source.
From the combined analysis of the available \xmm\ and
\chan\ data, we determine a coronal temperature of about $1$~MK and and emission
measure of $7\,(4-9) \times 10^{49}$~cm$^{-3}$. 

With respect to coronal temperature, \hd\ appears to be comparable to the
inactive Sun or \acen~A. We find that, in terms of HRC-I count
rate, \hd\ is more luminous
than \acen~A in its low activity state. However, \acen~A and \hd\ may
of course both vary in their luminosity, possibly due to cycles, which
can only be studied with long-term monitoring programs. \hd\ rotates faster
than the Sun and \acen~A, which favors higher activity levels.
This is also consistent with the distance-scaled HRC-I count rate of the
Maunder-Minimum candidate 51~Peg being lower than that of \hd.
It appears, however, that \hd\ is a faint X-ray emitter
with respect to its rotation period of $14.4$~d. The relation
between period and X-ray luminosity by \citet{Pizzolato2003}, for instance,
predicts an X-ray luminosity more than an order of magnitude
higher than what we observe. Whether this bears some deeper meaning,
is merely a coincidence, or a consequence of the relations being uncertain in
this regime cannot be decided with the data at hand.

In short, \hd\ is an inactive star in coronal X-rays,
comparable to the Sun itself.
The coronal properties of \hd\
derived in our analysis are consistent with those used by \citet{Louden2016} in their study and thus,
demonstrate that the level of planetary atmospheric irradiation is sufficient to
drive planetary evaporation at a rate of a few times $10^{10}$~g\,s$^{-1}$.

\begin{acknowledgements}
We thank the anonymous referee for a knowledgeable and helpful report.
This work has made use of data from the European Space Agency (ESA)
mission {\it Gaia} (\url{http://www.cosmos.esa.int/gaia}), processed by
the {\it Gaia} Data Processing and Analysis Consortium (DPAC,
\url{http://www.cosmos.esa.int/web/gaia/dpac/consortium}). Funding
for the DPAC has been provided by national institutions, in particular
the institutions participating in the {\it Gaia} Multilateral Agreement.
SC and MM acknowledge support by UHH. CS and MS acknowledge funding through DFG.
\end{acknowledgements} 

\bibliographystyle{aa}
\bibliography{doc.bib}

\begin{thebibliography}{46}
\expandafter\ifx\csname natexlab\endcsname\relax\def\natexlab#1{#1}\fi

\bibitem[{{Arnaud}(1996)}]{Arnaud1996}
{Arnaud}, K.~A. 1996, in Astronomical Society of the Pacific Conference Series,
  Vol. 101, Astronomical Data Analysis Software and Systems V, ed. G.~H.
  {Jacoby} \& J.~{Barnes}, 17

\bibitem[{{Ayres}(2014)}]{Ayres2014}
{Ayres}, T.~R. 2014, \aj, 147, 59

\bibitem[{{Baliunas} {et~al.}(1995){Baliunas}, {Donahue}, {Soon}, {Horne},
  {Frazer}, {Woodard-Eklund}, {Bradford}, {Rao}, {Wilson}, {Zhang}, {Bennett},
  {Briggs}, {Carroll}, {Duncan}, {Figueroa}, {Lanning}, {Misch}, {Mueller},
  {Noyes}, {Poppe}, {Porter}, {Robinson}, {Russell}, {Shelton}, {Soyumer},
  {Vaughan}, \& {Whitney}}]{Baliunas1995}
{Baliunas}, S.~L., {Donahue}, R.~A., {Soon}, W.~H., {et~al.} 1995, \apj, 438,
  269

\bibitem[{{Bazot} {et~al.}(2007){Bazot}, {Bouchy}, {Kjeldsen}, {Charpinet},
  {Laymand}, \& {Vauclair}}]{Bazot2007}
{Bazot}, M., {Bouchy}, F., {Kjeldsen}, H., {et~al.} 2007, \aap, 470, 295

\bibitem[{{Bonfanti} {et~al.}(2016){Bonfanti}, {Ortolani}, \&
  {Nascimbeni}}]{Bonfanti2016}
{Bonfanti}, A., {Ortolani}, S., \& {Nascimbeni}, V. 2016, \aap, 585, A5

\bibitem[{{Boyajian} {et~al.}(2015){Boyajian}, {von Braun}, {Feiden}, {Huber},
  {Basu}, {Demarque}, {Fischer}, {Schaefer}, {Mann}, {White}, {Maestro},
  {Brewer}, {Lamell}, {Spada}, {L{\'o}pez-Morales}, {Ireland}, {Farrington},
  {van Belle}, {Kane}, {Jones}, {ten Brummelaar}, {Ciardi}, {McAlister},
  {Ridgway}, {Goldfinger}, {Turner}, \& {Sturmann}}]{Boyajian2015}
{Boyajian}, T., {von Braun}, K., {Feiden}, G.~A., {et~al.} 2015, \mnras, 447,
  846

\bibitem[{{Charbonneau} {et~al.}(2000){Charbonneau}, {Brown}, {Latham}, \&
  {Mayor}}]{Charbonneau2000}
{Charbonneau}, D., {Brown}, T.~M., {Latham}, D.~W., \& {Mayor}, M. 2000, \apjl,
  529, L45

\bibitem[{{Daemgen} {et~al.}(2009){Daemgen}, {Hormuth}, {Brandner}, {Bergfors},
  {Janson}, {Hippler}, \& {Henning}}]{Daemgen2009}
{Daemgen}, S., {Hormuth}, F., {Brandner}, W., {et~al.} 2009, \aap, 498, 567

\bibitem[{{del Burgo} \& {Allende Prieto}(2016)}]{delBurgo2016}
{del Burgo}, C. \& {Allende Prieto}, C. 2016, \mnras, 463, 1400

\bibitem[{{Eggenberger} {et~al.}(2004){Eggenberger}, {Charbonnel}, {Talon},
  {Meynet}, {Maeder}, {Carrier}, \& {Bourban}}]{Eggenberger2004}
{Eggenberger}, P., {Charbonnel}, C., {Talon}, S., {et~al.} 2004, \aap, 417, 235

\bibitem[{Foster {et~al.}(2012)Foster, Ji, Smith, \& Brickhouse}]{Foster2012}
Foster, A.~R., Ji, L., Smith, R.~K., \& Brickhouse, N.~S. 2012, The
  Astrophysical Journal, 756, 128

\bibitem[{{Gaia Collaboration} {et~al.}(2016{\natexlab{a}}){Gaia
  Collaboration}, {Brown}, {Vallenari}, {Prusti}, {de Bruijne}, {Mignard},
  {Drimmel}, {Babusiaux}, {Bailer-Jones}, {Bastian}, \& et~al.}]{Gaia2016a}
{Gaia Collaboration}, {Brown}, A.~G.~A., {Vallenari}, A., {et~al.}
  2016{\natexlab{a}}, \aap, 595, A2

\bibitem[{{Gaia Collaboration} {et~al.}(2016{\natexlab{b}}){Gaia
  Collaboration}, {Prusti}, {de Bruijne}, {Brown}, {Vallenari}, {Babusiaux},
  {Bailer-Jones}, {Bastian}, {Biermann}, {Evans}, \& et~al.}]{Gaia2016b}
{Gaia Collaboration}, {Prusti}, T., {de Bruijne}, J.~H.~J., {et~al.}
  2016{\natexlab{b}}, \aap, 595, A1

\bibitem[{{Henry} {et~al.}(2000){Henry}, {Marcy}, {Butler}, \&
  {Vogt}}]{Henry2000}
{Henry}, G.~W., {Marcy}, G.~W., {Butler}, R.~P., \& {Vogt}, S.~S. 2000, \apjl,
  529, L41

\bibitem[{{Isaacson} \& {Fischer}(2010)}]{Isaacson2010}
{Isaacson}, H. \& {Fischer}, D. 2010, \apj, 725, 875

\bibitem[{Jaynes(1968)}]{Jaynes1968}
Jaynes, E.~T. 1968, IEEE Transactions on Systems Science and Cybernetics, 4,
  227

\bibitem[{{Kervella} {et~al.}(2003){Kervella}, {Th{\'e}venin}, {S{\'e}gransan},
  {Berthomieu}, {Lopez}, {Morel}, \& {Provost}}]{Kervella2003}
{Kervella}, P., {Th{\'e}venin}, F., {S{\'e}gransan}, D., {et~al.} 2003, \aap,
  404, 1087

\bibitem[{{Knutson} {et~al.}(2010){Knutson}, {Howard}, \&
  {Isaacson}}]{Knutson2010}
{Knutson}, H.~A., {Howard}, A.~W., \& {Isaacson}, H. 2010, \apj, 720, 1569

\bibitem[{{Louden} {et~al.}(2016){Louden}, {Wheatley}, \&
  {Briggs}}]{Louden2016}
{Louden}, T., {Wheatley}, P.~J., \& {Briggs}, K. 2016, ArXiv e-prints
  [\eprint[arXiv]{1605.07987}]

\bibitem[{{Mamajek} \& {Hillenbrand}(2008)}]{Mamajek2008}
{Mamajek}, E.~E. \& {Hillenbrand}, L.~A. 2008, \apj, 687, 1264

\bibitem[{{Maxted} {et~al.}(2015){Maxted}, {Serenelli}, \&
  {Southworth}}]{Maxted2015}
{Maxted}, P.~F.~L., {Serenelli}, A.~M., \& {Southworth}, J. 2015, \aap, 577,
  A90

\bibitem[{{Mayor} \& {Queloz}(1995)}]{Mayor1995}
{Mayor}, M. \& {Queloz}, D. 1995, \nat, 378, 355

\bibitem[{{Mittag} {et~al.}(2011){Mittag}, {Hempelmann},
  {Gonz{\'a}lez-P{\'e}rez}, {Schmitt}, \& {Hall}}]{Mittag2011}
{Mittag}, M., {Hempelmann}, A., {Gonz{\'a}lez-P{\'e}rez}, J.~N., {Schmitt},
  J.~H.~M.~M., \& {Hall}, J.~C. 2011, in Astronomical Society of the Pacific
  Conference Series, Vol. 448, 16th Cambridge Workshop on Cool Stars, Stellar
  Systems, and the Sun, ed. C.~{Johns-Krull}, M.~K. {Browning}, \& A.~A.
  {West}, 1187

\bibitem[{{Mittag} {et~al.}(2016){Mittag}, {Schr{\"o}der}, {Hempelmann},
  {Gonz{\'a}lez-P{\'e}rez}, \& {Schmitt}}]{Mittag2016}
{Mittag}, M., {Schr{\"o}der}, K.-P., {Hempelmann}, A.,
  {Gonz{\'a}lez-P{\'e}rez}, J.~N., \& {Schmitt}, J.~H.~M.~M. 2016, \aap, 591,
  A89

\bibitem[{{Monet} {et~al.}(2003){Monet}, {Levine}, {Canzian}, {Ables}, {Bird},
  {Dahn}, {Guetter}, {Harris}, {Henden}, {Leggett}, {Levison}, {Luginbuhl},
  {Martini}, {Monet}, {Munn}, {Pier}, {Rhodes}, {Riepe}, {Sell}, {Stone},
  {Vrba}, {Walker}, {Westerhout}, {Brucato}, {Reid}, {Schoening}, {Hartley},
  {Read}, \& {Tritton}}]{Monet2003}
{Monet}, D.~G., {Levine}, S.~E., {Canzian}, B., {et~al.} 2003, \aj, 125, 984

\bibitem[{{Noyes} {et~al.}(1984){Noyes}, {Hartmann}, {Baliunas}, {Duncan}, \&
  {Vaughan}}]{Noyes1984}
{Noyes}, R.~W., {Hartmann}, L.~W., {Baliunas}, S.~L., {Duncan}, D.~K., \&
  {Vaughan}, A.~H. 1984, \apj, 279, 763

\bibitem[{{Peres} {et~al.}(2000){Peres}, {Orlando}, {Reale}, {Rosner}, \&
  {Hudson}}]{Peres2000}
{Peres}, G., {Orlando}, S., {Reale}, F., {Rosner}, R., \& {Hudson}, H. 2000,
  \apj, 528, 537

\bibitem[{{Pizzolato} {et~al.}(2003){Pizzolato}, {Maggio}, {Micela},
  {Sciortino}, \& {Ventura}}]{Pizzolato2003}
{Pizzolato}, N., {Maggio}, A., {Micela}, G., {Sciortino}, S., \& {Ventura}, P.
  2003, \aap, 397, 147

\bibitem[{{Poppenh{\"a}ger} {et~al.}(2009){Poppenh{\"a}ger}, {Robrade},
  {Schmitt}, \& {Hall}}]{Poppenhaeger2009}
{Poppenh{\"a}ger}, K., {Robrade}, J., {Schmitt}, J.~H.~M.~M., \& {Hall}, J.~C.
  2009, \aap, 508, 1417

\bibitem[{{Raghavan} {et~al.}(2006){Raghavan}, {Henry}, {Mason}, {Subasavage},
  {Jao}, {Beaulieu}, \& {Hambly}}]{Raghavan2006}
{Raghavan}, D., {Henry}, T.~J., {Mason}, B.~D., {et~al.} 2006, \apj, 646, 523

\bibitem[{{Robrade} {et~al.}(2012){Robrade}, {Schmitt}, \&
  {Favata}}]{Robrade2012}
{Robrade}, J., {Schmitt}, J.~H.~M.~M., \& {Favata}, F. 2012, \aap, 543, A84

\bibitem[{{Rosen} {et~al.}(2016){Rosen}, {Webb}, {Watson}, {Ballet}, {Barret},
  {Braito}, {Carrera}, {Ceballos}, {Coriat}, {Della Ceca}, {Denkinson},
  {Esquej}, {Farrell}, {Freyberg}, {Gris{\'e}}, {Guillout}, {Heil},
  {Koliopanos}, {Law-Green}, {Lamer}, {Lin}, {Martino}, {Michel}, {Motch},
  {Nebot Gomez-Moran}, {Page}, {Page}, {Page}, {Pakull}, {Pye}, {Read},
  {Rodriguez}, {Sakano}, {Saxton}, {Schwope}, {Scott}, {Sturm}, {Traulsen},
  {Yershov}, \& {Zolotukhin}}]{Rosen2016}
{Rosen}, S.~R., {Webb}, N.~A., {Watson}, M.~G., {et~al.} 2016, \aap, 590, A1

\bibitem[{{Salz} {et~al.}(2016{\natexlab{a}}){Salz}, {Czesla}, {Schneider}, \&
  {Schmitt}}]{Salz2016a}
{Salz}, M., {Czesla}, S., {Schneider}, P.~C., \& {Schmitt}, J.~H.~M.~M.
  2016{\natexlab{a}}, \aap, 586, A75

\bibitem[{{Salz} {et~al.}(2016{\natexlab{b}}){Salz}, {Schneider}, {Czesla}, \&
  {Schmitt}}]{Salz2016b}
{Salz}, M., {Schneider}, P.~C., {Czesla}, S., \& {Schmitt}, J.~H.~M.~M.
  2016{\natexlab{b}}, \aap, 585, L2

\bibitem[{{Santos} {et~al.}(2004){Santos}, {Israelian}, \&
  {Mayor}}]{Santos2004}
{Santos}, N.~C., {Israelian}, G., \& {Mayor}, M. 2004, \aap, 415, 1153

\bibitem[{{Sanz-Forcada} {et~al.}(2011){Sanz-Forcada}, {Micela}, {Ribas},
  {Pollock}, {Eiroa}, {Velasco}, {Solano}, \&
  {Garc{\'{\i}}a-{\'A}lvarez}}]{Sanz2011}
{Sanz-Forcada}, J., {Micela}, G., {Ribas}, I., {et~al.} 2011, \aap, 532, A6

\bibitem[{{Sanz-Forcada} {et~al.}(2010){Sanz-Forcada}, {Ribas}, {Micela},
  {Pollock}, {Garc{\'{\i}}a-{\'A}lvarez}, {Solano}, \& {Eiroa}}]{Sanz2010}
{Sanz-Forcada}, J., {Ribas}, I., {Micela}, G., {et~al.} 2010, \aap, 511, L8

\bibitem[{{Schmitt}(1997)}]{Schmitt1997}
{Schmitt}, J.~H.~M.~M. 1997, \aap, 318, 215

\bibitem[{{Schmitt} {et~al.}(2014){Schmitt}, {Schr{\"o}der}, {Rauw},
  {Hempelmann}, {Mittag}, {Gonz{\'a}lez-P{\'e}rez}, {Czesla}, {Wolter}, {Jack},
  {Eenens}, \& {Trinidad}}]{Schmitt2014}
{Schmitt}, J.~H.~M.~M., {Schr{\"o}der}, K.-P., {Rauw}, G., {et~al.} 2014,
  Astronomische Nachrichten, 335, 787

\bibitem[{{Silva-Valio}(2008)}]{SilvaValio2008}
{Silva-Valio}, A. 2008, \apjl, 683, L179

\bibitem[{{Torres} {et~al.}(2008){Torres}, {Winn}, \& {Holman}}]{Torres2008}
{Torres}, G., {Winn}, J.~N., \& {Holman}, M.~J. 2008, \apj, 677, 1324

\bibitem[{{Tsuneta} {et~al.}(1991){Tsuneta}, {Acton}, {Bruner}, {Lemen},
  {Brown}, {Caravalho}, {Catura}, {Freeland}, {Jurcevich}, {Morrison},
  {Ogawara}, {Hirayama}, \& {Owens}}]{Tsuneta1991}
{Tsuneta}, S., {Acton}, L., {Bruner}, M., {et~al.} 1991, \solphys, 136, 37

\bibitem[{{Vidal-Madjar} {et~al.}(2004){Vidal-Madjar}, {D{\'e}sert},
  {Lecavelier des Etangs}, {H{\'e}brard}, {Ballester}, {Ehrenreich}, {Ferlet},
  {McConnell}, {Mayor}, \& {Parkinson}}]{VidalMadjar2004}
{Vidal-Madjar}, A., {D{\'e}sert}, J.-M., {Lecavelier des Etangs}, A., {et~al.}
  2004, \apjl, 604, L69

\bibitem[{{Vidal-Madjar} {et~al.}(2003){Vidal-Madjar}, {Lecavelier des Etangs},
  {D{\'e}sert}, {Ballester}, {Ferlet}, {H{\'e}brard}, \&
  {Mayor}}]{VidalMadjar2003}
{Vidal-Madjar}, A., {Lecavelier des Etangs}, A., {D{\'e}sert}, J.-M., {et~al.}
  2003, \nat, 422, 143

\bibitem[{Winn {et~al.}(2005)Winn, Noyes, Holman, Charbonneau, Ohta, Taruya,
  Suto, Narita, Turner, Johnson, Marcy, Butler, \& Vogt}]{Winn2005}
Winn, J.~N., Noyes, R.~W., Holman, M.~J., {et~al.} 2005, The Astrophysical
  Journal, 631, 1215

\bibitem[{{Wood} {et~al.}(2005){Wood}, {Redfield}, {Linsky}, {M{\"u}ller}, \&
  {Zank}}]{Wood2005}
{Wood}, B.~E., {Redfield}, S., {Linsky}, J.~L., {M{\"u}ller}, H.-R., \& {Zank},
  G.~P. 2005, \apjs, 159, 118

\end{thebibliography}

\end{document}